\documentclass[superscriptaddress,amssymb,prb,10pt,twocolumn,showpacs]{revtex4-1}
\usepackage{amsmath}    
\pdfoutput=1
\usepackage{graphicx}   
\usepackage{verbatim}   
\usepackage{color}      
\usepackage{subfigure}
\usepackage{gensymb}  
\usepackage{textcomp}
\usepackage{longtable}
\usepackage{bm}
\usepackage{color}
\usepackage{epstopdf} 

\def\etal{{\it et~al.}}
\begin{document}
\title{\boldmath Structural, magnetic and superconducting properties of pulsed-laser-deposition-grown $\rm{La_{1.85}Sr_{0.15}CuO_{4}/La_{2/3}Ca_{1/3}MnO_{3}}$ superlattices on  $\rm{(001)}$-oriented $\rm{LaSrAlO_{4}}$ substrates \unboldmath}

\author{S. Das}
\email{saikat.das@unifr.ch}
\author{K. Sen}
\author{I. Marozau}
\author{ M. A. Uribe-Laverde}
\affiliation{University of Fribourg, Department of Physics and Fribourg Center for Nanomaterials,
Chemin du Mus\'{e}e 3, CH-1700 Fribourg, Switzerland}
\author{N. Biskup}
\affiliation {Departamento de F\'{i}sica Aplicada III and Instituto Pluridisciplinar, Universidad Complutense de Madrid, Spain.}
\author{M. Varela}
\affiliation {Departamento de F\'{i}sica Aplicada III and Instituto Pluridisciplinar, Universidad Complutense de Madrid, Spain.}
\affiliation {Materials Science and Technology Division, Oak Ridge National Laboratory, Oak Ridge, TN 37831, USA.}
\author{Y. Khaydukov}
\author{O. Soltwedel}
\author{T. Keller}
\affiliation{Max-Planck-Institut f\"{u}r Festk\"{o}rperforschung, Stuttgart, 70569, Germany}
\affiliation{Max Planck Society Outstation at the Forschungsneutronenquelle Heinz Maier-Leibnitz (FRM-II), D-85747 Garching, Germany}
\author{M. D\"{o}beli}
\affiliation{ Laboratory of Ion Beam Physics, ETH Zurich, Schafmattstrasse 20, CH-8093 Zurich, Switzerland}
\author{C. W. Schneider}
\affiliation{Paul Scherrer Institut, CH-5232 Villigen, Switzerland}
\author{C. Bernhard}
\email{christian.bernhard@unifr.ch}
\affiliation{University of Fribourg, Department of Physics and Fribourg Center for Nanomaterials,
Chemin du Mus\'{e}e 3, CH-1700 Fribourg, Switzerland}

\begin{abstract}
Epitaxial $\rm{La_{1.85}Sr_{0.15}CuO_{4}/La_{2/3}Ca_{1/3}MnO_{3}}$ (LSCO/LCMO) superlattices (SL) on $\rm{(001)}$-oriented $\rm{LaSrAlO_{4}}$ substrates have been grown with pulsed laser deposition (PLD) technique. Their  structural, magnetic and superconducting properties have been determined with \textit{in-situ} reflection high energy electron diffraction (RHEED), x-ray diffraction, specular neutron reflectometry, scanning transmission electron microscopy (STEM), electric transport, and magnetization measurements. 
We find that despite the large mismatch between the in-plane lattice parameters of LSCO (a = 0.3779 nm) and LCMO (a = 0.387 nm) these superlattices can be grown epitaxially and with a high crystalline quality. While the first LSCO layer remains clamped to the LSAO substrate, a sizeable strain relaxation occurs already in the first LCMO layer. The following LSCO and LCMO layers adopt a nearly balanced state in which the tensile and compressive strain effects yield alternating in-plane lattice parameters with an almost constant average value. No major defects are observed in the LSCO layers, while a significant number of vertical antiphase boundaries are found in the LCMO layers. The LSCO layers remain superconducting with a relatively high superconducting onset temperature of $\rm{T^{onset}_{c}\approx36}$ K. The macroscopic superconducting response is also evident in the magnetization data due to a weak diamagnetic signal below 10 K for H $\parallel$ ab and a sizeable paramagnetic shift for H $\parallel$ c that can be explained in terms of a vortex-pinning-induced flux compression. The LCMO layers maintain a strongly ferromagnetic state with a Curie temperature of $\rm{T^{Curie}\approx190}$ K and a large low-temperature saturation moment of about 3.5(1) $\rm{\mu_{B}}$. These results suggest that the LSCO/LCMO superlattices can be used to study the interaction between the antagonistic ferromagnetic and superconducting orders and, in combination with previous studies on YBCO/LCMO superlattices, may allow one to identify the relevant mechanisms. 
\end{abstract}


\maketitle

\section{Introduction}
Artificially-grown multilayers from cuprate high temperature superconductors (SC) and ferromagnetic (FM) manganites are unique model systems to study the interplay between the antagonistic superconducting and ferromagnetic order parameters over a wide range of temperatures and magnetic fields. A number of fascinating phenomena have already been discovered in $\rm{YBa_{2}Cu_{3}O_{7-x}/La_{2/3}Ca_{1/3}MnO_{3}}$ (YBCO/LCMO) heterostructures such as a large photo-induced enhancement of the superconducting transition temperature, $\rm{T_{c}}$ \cite{Pena2006}, a giant magnetoresistance effect \cite{Pena2005}, or an unusual superconductivity-induced modulation of the ferromagnetic order \cite{Hoppler2009}. Another prominent example is the magnetic proximity effect (MPE) which gives rise to a strong suppression of the FM moment of the Mn ions on the LCMO side of the interface and yet a small induced FM moment of the Cu ions (that is antiparallel to the one of Mn) on the YBCO side \cite{Stahn2005,Chakhalian2006}. Recently, it has been shown that this MPE is strongly dependent on the doping state of the manganite layers, i.e. it is essentially absent in superlattices in which the ferromagnetic manganite layers have a reduced hole doping and thus remain insulating
\cite{Satapathy2012}. This observation calls for a corresponding study of the influence of the hole doping state of the cuprate layers on the MPE. For the YBCO/LCMO multilayers this requires a controlled variation of the oxygen content of the CuO chains of YBCO layers which is relatively difficult to achieve for these thin film structures. In addition, transmission electron microscopy studies have shown that both the top and the bottom YBCO/LCMO interfaces exhibit equivalent stacking sequences of $\rm{CuO_{2}-Y-CuO_{2}-BaO-MnO_{2}}$ or vice versa. The $\rm{CuO_{2}}$  bilayer next to the interface is therefore always lacking one of its neighboring CuO chain layers and thus half of its charge reservoir. For fully oxygenated YBCO it has been shown that these missing CuO chain layers, and possibly an additional charge transfer across the interface between LCMO and YBCO \cite{Varela2003}, result in a weakly metallic and superconducting state of the $\rm{CuO_{2}}$ bilayers next to the interface. For the case of de-oxygenated YBCO these interfacial $\rm{CuO_{2}}$ bilayers would be strongly underdoped and thus likely remain insulating.

Some of these problems can be circumvented and new insight may be gained with corresponding cuprate/manganite multilayers in which YBCO is replaced with $\rm{La_{2-x}Sr_{x}CuO_{4}}$ (LSCO). LSCO has a simpler crystallographic structure and its doping state can be readily tuned via the Sr concentration, x, which can be varied starting from x = 0, where the system is a charge transfer insulator, through the optimally doped state at $\rm{x=0.15}$ with a maximum superconducting $\rm{T_{c}\approx 40}$ K \cite{Tarascon1987}, to the heavily overdoped and strongly metallic state. The use of LSCO instead of YBCO may also lead to a different interfacial layer stacking sequence and thus a different local bonding between the Cu and Mn ions at the interface. This could be a unique opportunity to check the universality of the MPE and to learn whether it is governed by the local bonding between Mn and Cu at the interface \cite{Chakhalian2007}.

A major challenge for the growth of high quality LSCO/LCMO superlattices concerns the sizeable mismatch of the in-plane lattice parameters of LSCO and LCMO which amounts to about -2.4 \% This is especially critical in view of the strong dependence of the superconducting transition temperature of LSCO thin films on strain effects due to the substrate \cite{Locquet1998}. It was also shown that the $\rm{T_{c}}$ values of thin films are extremely sensitive to the net oxygen content \cite{Bozovic2002}. The best quality LSCO thin films were grown either by ozone or atomic oxygen assisted molecular beam epitaxy (MBE) or by e-beam co-evaporation techniques \cite{Bozovic2002,Sato1997}. To date there exist only few reports of $\rm{La_{1.85}Sr_{0.15}CuO_{4}}$ films with bulk superconducting properties and high $\rm{T_{c}}$ values that have been grown with pulsed laser deposition (PLD) technique \cite{Trofimov1994,Si1999,Rakshit2004}. Those films have been deposited in molecular $\rm{O_{2}}$ environment and have either been oxygenated by slow cooling inside the growth chamber in a gas mixture of oxygen and ozone or by performing an \textit{ex-situ} post-growth annealing treatment in high pressure oxygen atmosphere. 

An alternative approach to achieve a stoichiometric oxygenation is to use a reactive gas atmosphere during the PLD growth, e.g. ozone ($\rm{O_{3}}$), $\rm{NO_{2}}$ or  $\rm{N_{2}O}$. However, $\rm{O_{3}}$ and $\rm{NO_{2}}$ are very corrosive gases which require a special design of the PLD chamber and its functional components that cannot be easily realized in most PLD systems. $\rm{N_{2}O}$ remains as an alternative since it is thermodynamically stable but delivers a higher concentration of atomic oxygen, via the collisions with energetic plasma particles during laser ablation, than $\rm{O_{2}}$ gas \cite{Gupta1993}. $\rm{N_{2}O}$ has already been successfully used for the PLD thin film growth of electron doped cuprates \cite{Kussmaul1992}. To the best of our knowledge it has not yet been used for the PLD growth of LSCO thin films. Nevertheless, it has recently been demonstrated that thin LCMO films grown with PLD in  $\rm{N_{2}O}$ atmosphere have a higher oxygen content and a superior crystalline quality than those grown in  $\rm{O_{2}}$ environment \cite{Esposito2011}. 

In this manuscript we describe the $\rm{N_{2}O}$ assisted PLD growth of $\rm{La_{1.85}Sr_{0.15}CuO_{4}}$ thin films and of  $\rm{La_{1.85}Sr_{0.15}CuO_{4}/La_{2/3}Ca_{1/3}MnO_{3}}$ superlattices. We also present a comprehensive study of their structural, magnetic, electronic, and superconducting properties. 

\section{Experiment}
Single LSCO thin films with a thickness of about $\rm{7.5-8}$ unit cells (u.c.) or $\rm{\approx10-10.6}$ nm and a series of [$\rm{La_{1.85}Sr_{0.15}CuO_{4}}$ (7.5 - 8 u.c.)/$\rm{La_{2/3}Ca_{1/3}MnO_{3}}$(23 u.c.)]$\rm{_{X}}$ superlattices with a repetition number of  X = 1, 3, 5, 7 and 9 bilayers (in the following we refer to them as X\_BL samples) have been grown in a PLD chamber (SURFACE-TEC GmbH), equipped with an excimer KrF laser source ($\rm{\lambda}$ = 248 nm, $\rm{t_{s}}$ = 25 ns) and an infrared laser (JENOPTIK, JOLD-$\rm{140}$-CAXF-$\rm{6}$A) for heating the substrate. Stoichiometric polycrystalline ceramics pellets of very high density (Surfacenet GmbH and Pi-Kem, 99.9 \% purity) were used as targets. Prior to every deposition the targets that are mounted on a rotatable carousel were preablated to condition the surface while the substrate was shielded from the plume with a shutter. For the deposition we used an on-axis ablation configuration with the substrate placed at a distance of $\rm{5}$ cm above the targets. During the growth the substrates were held at a fixed temperature of  730 $\rm{\degree C}$ that was controlled with an infrared pyrometer. The partial pressure of the background $\rm{N_{2}O}$ gas was set to $\rm{0.12}$ mbar. A homogeneous section of the laser beam was selected by a mask at the laser exit and imaged onto the target, resulting in a sharp rectangular spot with an area of  2.3 $\rm{ mm^{2}}$. The laser was operated at a repetition rate of $\rm{2}$ Hz and a fluency of 1.2 $\rm{ J/cm^{2}}$ and 1.8 $\rm{J/cm^{2}}$ was used to grow the LSCO and LCMO layers, respectively. Right after the deposition the  $\rm{N_{2}O}$ gas flow was stopped and a flow of pure oxygen gas with a pressure of about $\rm{0.3}$ mbar in the chamber was used to flush out residual  $\rm{N_{2}O}$. The samples were annealed at this condition for about $30$ minutes before the chamber was vented with 1 bar of oxygen and the samples were cooled to 550 $\rm{\degree C}$  at a rate of  5 $\rm{\degree C}$/min where they were further annealed for 1 hour. Finally, the samples were rapidly cooled to room temperature with a rate of  30 $\rm{\degree C}$/min. For most samples no additional \textit{ex-situ} annealing was performed. 

The LSCO thin film and the LSCO/LCMO superlattices were grown on single crystalline $\rm{(001)}$-oriented $\rm{LaSrAlO_{4}}$ (LSAO) substrates with an area (5x5 $\rm{mm^{2}}$) (from MTI Corp, USA) \cite{MTICorp}. LSAO crystallizes in a perovskite-like tetragonal $\rm{K_{2}NiF_{4}}$ structure with room temperature lattice parameters a = b = 0.3756 nm and c = 1.263 nm \cite{Cieplak1994}. It is rather well lattice-matched with LSCO with lattice parameters a = b = 0.3777 nm and c = 1.323 nm \cite{Takagi1989} for which it yields a weak in-plane bi-axial compressive strain of about -0.6 \%. The in-plane lattice mismatch between pseudo-cubic LCMO with a = b = c = 0.387 nm \cite{Huang1998} and $\rm{LaSrAlO_{4}}$ is considerably larger and amounts to about -3 \%. Prior to the deposition each substrate was rinsed and cleaned in acetone and iso-propanol. A specific treatment to obtain a particular type of surface termination, similar as it was recently reported \cite{Biswas2013}, has not been performed.

The surface morphology during growth was routinely monitored with \textit{in-situ} Reflection High Energy Electron Diffraction (RHEED). A collimated beam of $\rm{30}$ keV electrons (R-DEC Co. Ltd., Japan) was directed at a glancing angle to the substrate surface along  $[110]$ crystallographic direction. The resulting diffraction pattern was recorded on a phosphorus screen. The RHEED data acquisition and subsequent analysis were carried out using the kSA $\rm{400}$ software. The grazing angle of incidence geometry makes RHEED an extremely surface sensitive tool which enables one to track and control the thin film growth mode as to realize atomically smooth surfaces and interfaces. For a layer-by-layer growth mode RHEED allows a precise control of the deposition with a resolution on the sub-unit cell level \cite{Ichimiya}.

The film surface morphology was examined by atomic force microscopy (AFM) at room temperature under ambient conditions with an NT-MDT NTEGRA Aura microscope.

The stoichiometric composition of the samples was analyzed  by Rutherford backscattering spectroscopy (RBS). This technique has a high sensitivity to heavier elements like La, Mn,Cu and in particular, allows for a rather accurate determination of their concentration ratio. It is unfortunately not very sensitive to lighter elements like oxygen. LSCO and LCMO thin films of thickness $\approx$ 100 nm were deposited on MgO (0 0 1) substrates under the growth conditions reported above. Mg is lighter than the La, Ca, Sr, Mn and Cu ions of the films. This strongly reduces  the overlap between the signals from the film and the substrate which helps to improve the compositional assessment.  The RBS spectra were collected using a 2 MeV $^{4}$He ion beam and a silicon surface barrier detector that was held at an angle of $\rm{168\degree}$. The data were simulated by the RUMP software. The experimental uncertainty in the ratio of the cation concentration is about $\pm$ 3 \%. 

The X-ray diffraction were performed with a four-circle diffractometer (Rigaku SmartLab) that is equipped with a $\rm{9}$ kW rotating anode Cu K$_{\alpha1}$ source, a parallel beam optics with a two-bounce Ge (220) monochromator $\rm{(\Delta\lambda/\lambda=3.8x10^{-4})}$ and a scintillation counter. The out-of-plane lattice parameters were obtained from symmetric 2Theta-Omega $\rm{(2\Theta-\omega)}$ scans over a wide angular range of  15\degree - 60\degree. The in-plane lattice parameters were derived from reciprocal space maps (RSM) around the (0 1 11) and (1 1 14) Bragg peaks of the LSAO substrate for which the (0 1 11) and (1 1 14) Bragg peaks of LSCO and the pseudo-cubic (0 1 3) and (1 1 4) Bragg peaks of LCMO are in close proximity. 
	
The room temperature specular unpolarized neutron reflectivity was measured at the NREX beamline at the FRM II neutron reactor in Munich, Germany. The beamline is equipped with an angle dispersive fixed-wavelength reflectometer which employs a continuous beam of monochromatic ($\rm{\lambda=0.426}$ nm) neutrons with a wavelength resolution of $\rm{\Delta\lambda/\lambda=\text{1-2 \%}}$ \cite{NREX}. The sample was mounted inside a closed cycle cryostat, the incident beam-width was set to 1 mm to ensure a full illumination of the sample surface (width = 5 mm) and yet a small background count. The reflectivity profile was fitted using the GenX software \cite{GenX}.

The cross-sectional high resolution scanning transmission electron microscopy (STEM) images of the SL were taken using an aberration corrected JEOL JEM-ARM200 CF, operated at 200 kV and equipped with a Gatan Quantum electron energy loss spectrometer. All images presented here are obtained using high angle annular dark field (ADF) imaging, also known as Z-contrast imaging. For this technique, the scattering cross section is given by Rutherford's law, i.e. the intensity of every atomic column is roughly proportional to the square of the atomic number Z. The contrast associated with heavier elements, such as La or Sr, is brighter, while lighter heavy elements, such as Cu or Mn, appear darker. The O atoms, being light and close to the heavier columns, are usually not visible in the ADF images. The specimens were prepared by conventional methods of grinding and Ar ion milling.
	
The electrical transport and dc-magnetization measurements were performed with a Quantum Design PPMS 9T system. The temperature dependent resistance was measured by attaching four wires at the corners of the samples using a bridge excitation current of 1000 $\rm{\mu}$A. A cooling rate of 2 K/min was used. The dc-magnetization data were obtained with the vibrating sample magnetometer (VSM) option. For the measurements with the field parallel to the film surface the samples were glued on a quartz holder. The corresponding out-of plane magnetic measurements with the field perpendicular to the surface (along the c-axis of LSCO) were carried out by placing the samples in a custom-made teflon holder. The magnetic response of the sample holders was carefully characterized and calibrated to exclude any residual paramagnetic or ferromagnetic signal. The temperature dependence of the magnetic moment was recorded during warming with a ramping rate of 2 K/min. The M-H loops at 10 K were recorded by sweeping the magnetic field at a rate of 11 Oe/sec. The as-measured data were corrected for the diamagnetic contribution of the substrate as to extract the intrinsic magnetic response of the thin film samples.

\section{Results and Discussions}
\subsection{{RHEED}}
\label{RHEED_sec}
\begin{figure*}
\centering
\vspace*{0cm}
\includegraphics[scale=.9]{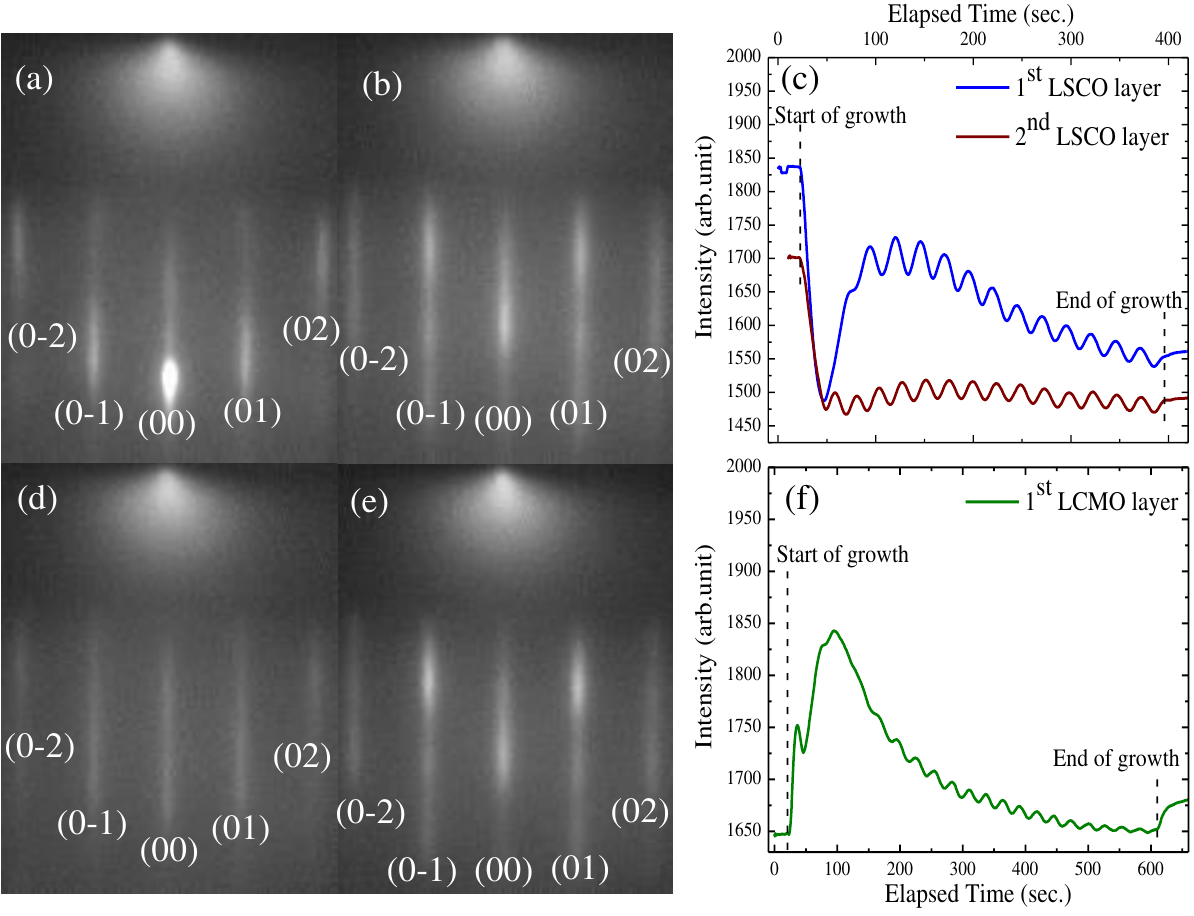}
\vspace*{0cm}
\caption{\label{RHEED} (Color online) Images of the RHEED pattern of (a) the  $\rm{1^{st}}$ LSCO layer, (b) the  $\rm{1^{st}}$ LCMO layer, (d) the  $\rm{2^{nd}}$ LSCO layer and (e) the last LCMO layer of the SL. The time evolution of the average intensity of the (00) Bragg peak during the growth is shown in (c) for the  $\rm{1^{st}}$ LSCO layer (blue line, top) and  for the  $\rm{2^{nd}}$ LSCO layer (brown line, bottom), and in (f) for the $\rm{1^{st}}$ LCMO layer.}
\end{figure*}
Figure \ref{RHEED} shows representative RHEED patterns obtained at different stages during the PLD growth of a nominal [LSCO (7.5 u.c.)/LCMO (23 u.c.)]$\rm{_{9}}$ SL. Very similar RHEED patterns were obtained for the other SLs and the single LSCO thin film. Figures \ref{RHEED}(c) and (f) show that the temporal evolution of the intensity of the specular (00) Bragg reflex exhibits an oscillatory behavior during the growth of both the LSCO and the LCMO layers. This is a characteristic signature of a Frank-van der Merwe (layer-by-layer) growth mode. The oscillations are a consequence of the periodic roughening and smoothening of the surface of the growing film that occurs as the adatoms arrive from the plasma plume and crystallize to form a new monolayer. For the case of LSCO one oscillation marks the growth of two LaO and one $\rm{CuO_{2}}$ atomic planes, this corresponds to half a crystallographic unit cell with a thickness of about 0.66 nm \cite{Terashima1990}. For the LCMO perovskite structure each oscillation marks the growth of  one pseudo-cubic unit cell with a thickness of about 0.387 nm. The RHEED pattern of the $\rm{1^{st}}$ LSCO layer in Figure \ref{RHEED}(a) consists of well-defined 2D diffraction spots and streaks for which the intensity maxima are lying on a semicircular arc ($\rm{0^{th}}$ order Laue's ring). Such a pattern is characteristic of an atomically smooth surface that consists of flat and large two-dimensional ($2$D) islands.
Figure \ref{RHEED}(c) depicts the typical time dependence of the RHEED signal during the growth of the  $\rm{1^{st}}$ LSCO layer on top of the LSAO substrate (blue curve) and of the $\rm{2^{nd}}$ LSCO layer on top of the $\rm{1^{st}}$ LCMO layer (brown curve), respectively. Both curves exhibit pronounced growth oscillations from which an average growth rate of 0.026 nm/sec is extracted. The only exception concerns the very $\rm{1^{st}}$ LSCO unit cell grown directly on top of the LSAO substrate for which no growth oscillations are discernible. This is suggestive of a difference in the growth dynamics of this first LSCO unit cell which may be related to the mixed-termination of the LSAO surface layer consisting of random patches of La/Sr-O and $\rm{AlO_{2}}$. \begin{figure*}
\centering
\vspace*{0cm}
\hspace*{.5cm}
\includegraphics[scale=1]{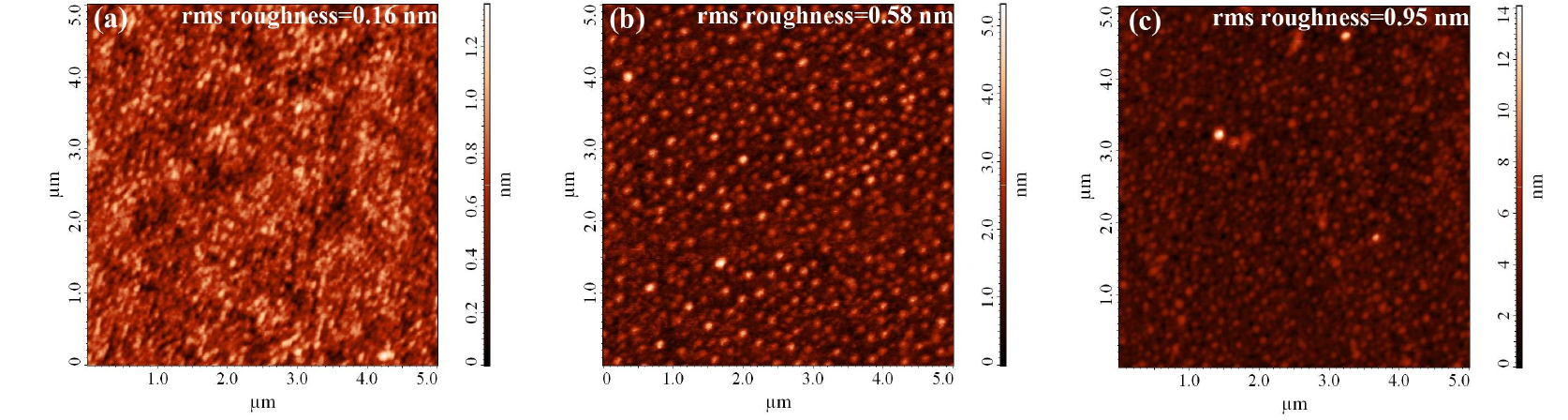} 
\vspace*{0cm}
\caption{\label{AFM} (Color online) AFM images of (a) the single LSCO thin film, (b) the 1\_BL sample and (c) the 9\_BL sample. The corresponding rms roughness is shown in the upper right corner of each image.}
\end{figure*}

Next we turn to the growth of the $\rm{1^{st}}$ LCMO layer. Figure \ref{RHEED}(b) shows that the specular Bragg streaks become more elongated along the vertical direction and their intensity is modulated such that new maxima appear at positions that do not follow the $\rm{0^{th}}$ order Laue's ring. This is an indication for the nucleation of some $3$D crystallites and a decrease of the correlation length of the $2$D islands. On the other hand, Figure \ref{RHEED}(f) confirms that the temporal oscillation of the specular RHEED intensity persists to the end of the growth of the LCMO layer. The oscillation period yields an average LCMO growth rate of $0.014$ nm/sec. The overall intensity of the $(00)$ Bragg peak increases at first, but then saturates and starts to decrease for a thickness of more than $3$-$4$ LCMO unit cells. After the end of the growth the intensity shows a partial recovery caused by slight smoothening of the surface due to rearrangement of mobile adatoms. The decay of the intensity profile is typical when the growth front gradually develops a larger roughness. This gradual change in the surface morphology is caused by the sizeable mismatch between the in-plane lattice parameters of LSCO and LCMO. As the growth of the LCMO layer proceeds, this increases the surface energy and thus favors the cohesion of adatoms among themselves rather than their adhesion to the growth front. 

In the RHEED data of the $\rm{2^{nd}}$ LSCO layer  Fig. \ref{RHEED}(d) there is no sign of a modulation of the streak intensity with peaks outside the $\rm{0^{th}}$ order Laue's ring (as was observed for the $\rm{1^{st}}$ LCMO layer). However, the streaks are less intense and vertically more elongated. This in association with the persistent oscillations recorded during the growth  Fig. \ref{RHEED}(c) (brown curve) indicates a purely 2D growth mode, but with a significantly reduced size of the 2D islands as compared to the $\rm{1^{st}}$ LSCO layer. This difference may be understood in terms of the lattice mismatch between the $\rm{2^{nd}}$ LSCO layer and the  $\rm{1^{st}}$ LCMO layer which is partially relaxed and thus has an increased in-plane lattice parameter. Yet, the RHEED data suggest that the strain effects are less disturbing for the growth mode of the $\rm{2^{nd}}$ LSCO layer than for the  $\rm{1^{st}}$ LCMO layer. A quantitative analysis of the RHEED data, to directly monitor the changes of the lattice parameters, is unfortunately not possible due to the broadening of the RHEED signal by the scattering of the electrons from the high pressure background gas. However, our interpretation that the LCMO layers are more strongly affected by the strain effects than the LSCO layers is supported by the x-ray and the STEM data as will be discussed in the following paragraphs.

Finally, we note that for the subsequent LCMO and LSCO layers, the RHEED data reveal similar trends as the ones reported above for the $\rm{1^{st}}$ LCMO and the$\rm{2^{nd}}$ LSCO layer, respectively. This is demonstrated in Fig. \ref{RHEED}(e) where the RHEED pattern of the $\rm{9^{th}}$ (and thus last) LCMO layer is still relatively sharp and characteristic of a fairly smooth surface. 

To summarize, the \textit{in-situ} RHEED data show that all the LSCO and LCMO layer have a predominantly two dimensional layer-by-layer type growth mode. Only in the LCMO layers we find some indication for the formation of some $3$D crystallites that seem to be caused by the large strain effects and related defects. 
\subsection{AFM}
\label{AFM_sec}
To study the surface morphology in real space, we performed atomic force microscopy (AFM) scans. Figures \ref{AFM}(a)-(c) show such large area scans (5x5 $\rm{\mu m^{2}}$) and the calculated values of the root mean square (rms) roughness for the single LSCO thin film and the 1\_BL and 9\_BL samples, respectively. The AFM image of the thin LSCO film reveals an atomically flat surface topography that is free of particles. The very small roughness of about 0.16 nm compares well with the sharp RHEED pattern observed during the growth of the $\rm{1^{st}}$ LSCO layer on LSAO (see Fig. \ref{RHEED}(a)). The surface of the 1\_BL sample contains some homogenously distributed, small particles which results in an enhanced value of the rms roughness of about 0.58 nm. This is again consistent with our interpretation of the modulated RHEED streak pattern during the growth of the $\rm{1^{st}}$ LCMO layer in Fig. \ref{RHEED}(b). Finally, the AFM micrograph for the 9\_BL  sample shows a few bigger particles together with an ensemble of homogeneously distributed, small particles which yield an rms roughness of 0.95 nm. This result confirms the rather gradual increase of the surface roughness during the growth of the superlattice. Taking into account the total thickness of the 9\_BL sample of about $180$ nm, it argues for a fairly smooth topography of our superlattices. 

\subsection{X-ray diffraction}
\label{Xray_diffrac}
\begin{figure}
\centering
\vspace*{0cm}
\includegraphics[height=10cm,width=8.5cm]{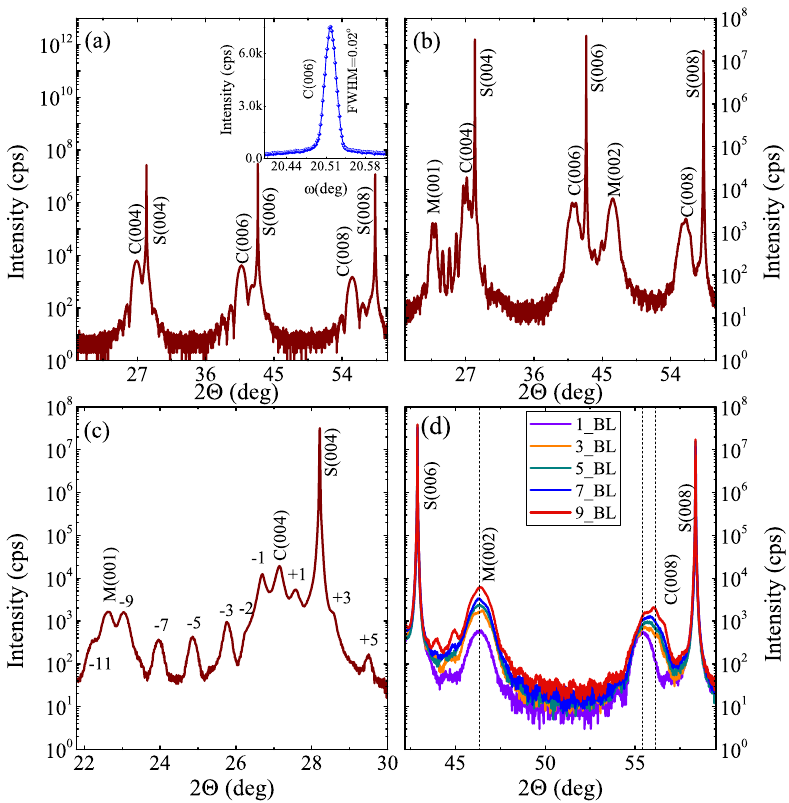}
\vspace*{0cm}
\caption{\label{SymXray} (Color online) Symmetric  $2\Theta-\omega$  scan of (a) the single LSCO thin film and (b) the 9\_BL sample. The inset of (a) shows a rocking scan around the (0 0 6) Bragg peak of LSCO. (c) Expanded view of the $2\Theta-\omega$ scan of the 9\_BL sample around the (0 0 4) Bragg peak of LSCO and the (0 0 1) peak of LCMO. The pronounced superlattice peaks are numbered according to their order. (d) Comparison of the $2\Theta-\omega$ scan of all the SLs showing the evolution of the LCMO (0 0 2) and LSCO (0 0 8) Bragg peaks. The vertical dashed lines mark the range over which LCMO $(0 0 2)$ and LSCO $(0 0 8)$ peaks are shifted.}
\end{figure}

Figure \ref{SymXray} shows representative x-ray diffraction curves for the symmetric $2\Theta-\omega$ scans of the single LSCO (8 u.c.) thin film and the LSCO (7.5-8 u.c.)/LCMO (23 u.c.) superlattices. The Bragg-peaks of LSAO, LSCO and LCMO are marked with the letters S, C and M, respectively. The scan for the single LSCO film in Fig. \ref{SymXray}(a) contains sharp (0 0 l) Bragg reflexes and is void of any peaks due to LSCO grains with a different orientation or from impurity phases. This confirms that the film is grown epitaxially with the c-axis pointing along the surface normal of the film. The high structural quality of the LSCO film is also documented by the rocking curve around the LSCO (0 0 6) Bragg peak, as shown in the inset, which is rather narrow with a full width at half maximum (FWHM) of about 0.02\degree . The average c-axis lattice parameter of the LSCO film, as extracted from the (0 0 4), (0 0 6) and (0 0 8) Bragg peaks, amounts to 1.326(2) nm. It is slightly larger than in bulk LSCO (c = 1.323 nm) as is expected for a non-relaxed film for which the in-plane lattice parameters is locked by the biaxial compressive strain of the LSAO substrate. The intensity exhibits a number of oscillations around the Bragg peaks of the LSCO film. These arise from the interference between the x-ray beams that are reflected from the top and the bottom interfaces of the LSCO film. These finite-size thickness oscillations or so-called Laue oscillations are indicative of a flat surface and a homogeneous thickness and density of the LSCO layer. The film thickness, as calculated from the period of these oscillations, amounts to 10.4 nm. This value is in fair agreement with the estimate from the RHEED oscillations which yields a thickness of 8 LSCO u.c. or about 10.6 nm.

Figure \ref{SymXray}(b) shows the symmetric  $2\Theta-\omega$ scan for the 9\_BL sample. It also exhibits pronounced (0 0 l) Bragg peaks of LSCO and pseudocubic LCMO and shows no sign of additional peaks due to misaligned grains or an impurity phase. This result is representative for all other SLs. Thanks to their different c-axis lattice parameters the (0 0 l) peaks for LSCO and LCMO are well separated such that they can be independently analyzed. Figure \ref{SymXray}(c) specifies the region around the (0 0 1) Bragg peak of LCMO and the (0 0 4) peak of LSCO. It reveals a series of sharp satellite peaks that are superimposed on the Bragg peaks of LCMO and LSCO. These so-called superlattice peaks arise from the constructive interference of the x-ray beams that are reflected from the top and the bottom interfaces of  the LSCO/LCMO bilayer units. They testify for the long-range coherent periodicity of these bilayers and for the high quality of the SL in terms of flat and homogeneous interfaces \cite{Schuller1980}. One can also notice that the odd-order superlattice peaks are very pronounced whereas the even-order superlattice peaks are barely visible. This confirms that the LSCO and LCMO layers have a very similar thickness. The thickness of the LSCO/LCMO bilayer as deduced from the oscillation period amounts to 19.9 nm. It agrees well with the estimate from the RHEED analysis as presented in \ref{RHEED_sec}.

Figure \ref{SymXray}(d) compares the symmetric  $2\Theta-\omega$ scans for all the SLs in the vicinity of the (0 0 2) Bragg peak of LCMO and the (0 0 8) peak of LSCO. It shows how the c-axis lattice parameters of the LSCO and LCMO layers evolve as the number of the BL repetitions increases. It is apparent that the position of the pseudo-cubic LCMO (0 0 2) Bragg peak is hardly shifted. This implies that the different LCMO layers have very similar values of the c-axis lattice parameter. This is not the case for the (0 0 8) peak of LSCO which exhibits a significant shift toward higher $2\Theta$ angles as the number of bilayers increases. Notably, the largest shift in the peak position occurs between the 1\_BL and the 3\_BL samples. We note that similar shifts of the LSCO peak have been observed for the (0 0 4) and (0 0 6) peaks of the SLs. This shows that the $\rm{1^{st}}$ LSCO layer has a significantly large c-axis lattice parameter than the following LSCO layers. It suggests that the major part of the strain relaxation occurs in the $\rm{1^{st}}$ LCMO layer, whereas in the following LSCO/LCMO bilayers a nearly balanced state is reached between the tensile strain of the LSCO layers and the compressive strain of the LCMO layers. 
\begin{figure*}
\centering
\vspace*{0cm}
\includegraphics[scale=1.5]{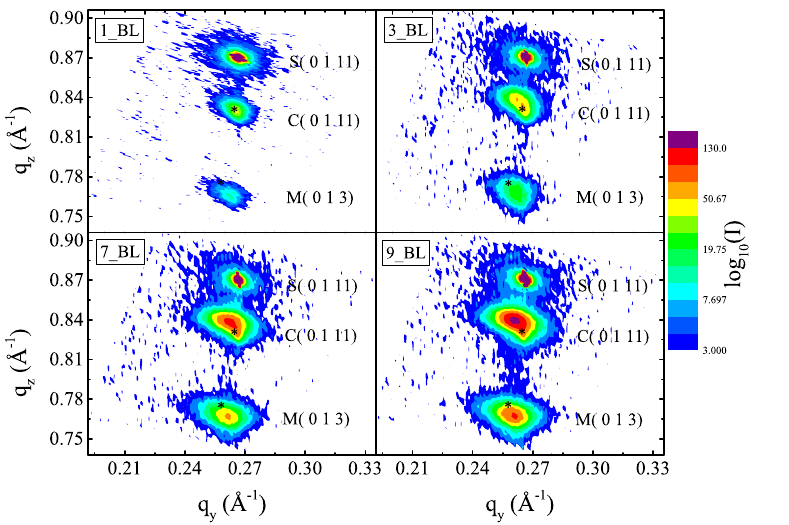}
\vspace*{0cm}
\caption{\label{RSM} (Color online) Reciprocal space maps around the (0 0 11) Bragg peak of the LSAO substrate for the 1\_BL, 3\_BL, 7\_BL and 9\_BL samples. The reciprocal lattice points of the bulk materials are indicated by the ($\ast$) symbols.}
\end{figure*}

 \begin{figure}
\centering
\vspace*{0cm}
\includegraphics[height=7cm]{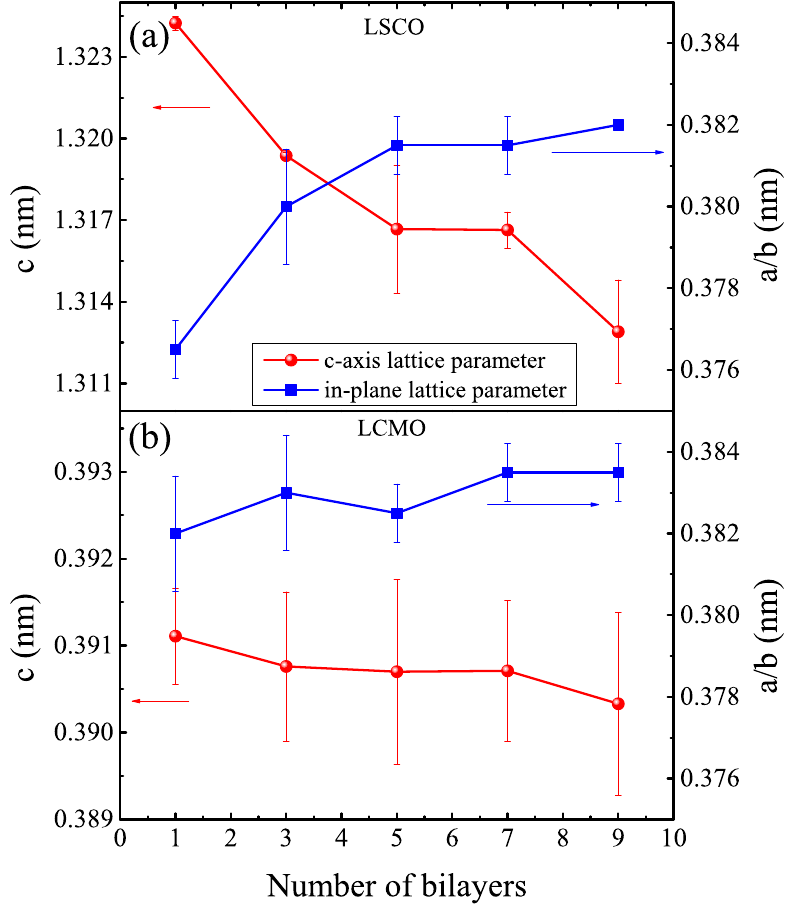}
\vspace*{0cm}
\caption{\label{LP} (Color online) Evolution of the lattice parameters as a function of the bilayer repetition for (a) the LSCO layers and (b) the LCMO layers. The average c-axis parameters have been extracted from the (0 0 4), (0 0 6) and (0 0 8) Bragg peaks for LSCO and from the (0 0 1) and (0 0 2) peaks for LCMO. The corresponding in-plane  parameters have been deduced from the Bragg peaks of the reciprocal space maps in the vicinity of the (0 1 11) and (1 1 14) peaks of LSAO. The error bars give the standard deviations. Solid lines are guides to the eye.}
\end{figure}

This interpretation is confirmed by the reciprocal space maps (RSM) which have been recorded around the asymmetric LSAO (0 1 11) and (1 1 14) Bragg peaks. For brevity, in Fig. \ref{RSM} we only show the reciprocal space maps around the (0 1 11) LSAO peak for the 1\_BL, 3\_BL, 7\_BL and 9\_BL samples. The asterisk symbols $(\ast)$ indicate the position of the corresponding Bragg peaks of bulk LSCO and LCMO. For the RSM of the  1\_BL sample the (0 1 11) peak of LSCO is collinear with the (0 1 11) peak of LSAO. This shows that the in-plane lattice parameter of the $\rm{1^{st}}$ LSCO layer matches the one of the LSAO substrate and suggests that the LSCO film is clamped to the LSAO substrate (which excerts a weak compressive strain). The pseudocubic (0 1 3) peak of the LCMO layer is clearly shifted toward lower $\rm{q_{y}}$ values, but it does not reach the position for bulk LCMO that is marked by the asterisk symbol. This shows that a sizeable, but still incomplete strain relaxation occurs in the $\rm{1^{st}}$ LCMO layer. The RSM of the  3\_BL sample in Fig. \ref{RSM}(b) reveals that the following LSCO layer undergoes a sizeable change of the in-plane as well as the c-axis lattice parameter. To a large extent, the in-plane lattice parameter approaches the one of the underlying LCMO layer, i.e. it increases and becomes somewhat larger than in bulk LSCO. For the SLs with an even larger numbers of BL units, the Bragg peaks of LSCO and LCMO broaden significantly along the vertical and the horizontal directions. The position of the peak intensity, corresponding to the average in-plane and c-axis lattice parameters, can still be reliably obtained by fitting the line-cuts with Gaussian profiles. The evolution of the c-axis and the in-plane lattice parameters of LSCO and LCMO, as obtained from the $2\Theta-\omega$ scans and the reciprocal space maps, is shown in Fig. \ref{LP}. It confirms that the LSCO layers exhibit a significantly larger variation of the in-plane and the c-axis lattice parameters than the LCMO layers. Notably, the biggest difference occurs between the $\rm{1^{st}}$ LSCO layer and the following ones. In comparison the in-plane and the c-axis lattice parameters of the LCMO layers undergo fairly moderate changes. All together the X-ray diffraction data suggest that the major part of the strain relaxation occurs within the $\rm{1^{st}}$ LCMO layer. This strain relaxation is incomplete, as compared to the bulk values of LCMO, and in the following LCMO and LSCO layers a nearly balanced state is acquired in which the in-plane lattice parameters of LSCO and LCMO alternate around an average value that changes only slowly. These LSCO layers are under a sizeable tensile strain and the LCMO layers under a corresponding compressive strain. Finally we mention that the calculated cell volumes for the LSCO and LCMO unit cells are quite similar for all the SL and deviate by less than $1.6\%$ from their bulk values. 
\subsection{Specular neutron reflectometry}
\label{Spec_Neutron}
\begin{figure}
\centering
\vspace*{0cm}
\includegraphics[scale=1]{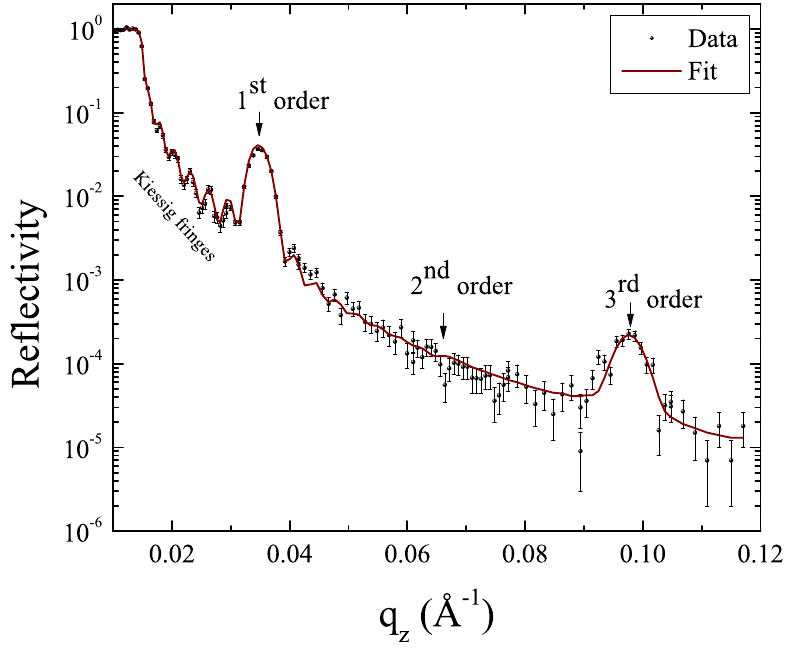}
\vspace*{0cm}
\caption{\label{Neutron} (Color online) Unpolarized neutron reflectivity curve of a LSCO/LCMO [LSCO (7.5 u.c.)/LCMO (25 u.c.)]$\rm{_{9}}$ SL measured at room temperature. Data are shown by symbols, the best fit by the solid line.}
\end{figure}

Neutron reflectometry is a well suited technique to probe on a truly macroscopic lateral length scale the quality of the interfaces and the homogeneity of the layer thicknesses of thin films and multilayer samples. This is especially true for the LSCO/LCMO SLs for which the difference in the nuclear scattering length density (SLD) of LSCO and LCMO is rather large (the difference in the x-ray SLD is very small). 
Figure \ref{Neutron} displays the room temperature specular neutron reflectivity curve of the [LSCO (7.5 u.c.)/LCMO (25 u.c.)]$\rm{_{9}}$ SL. It exhibits characteristic features, like Kiessig-fringes and sharp superlattice Bragg peaks that are indicative of a high structural quality. The pronounced Kiessig fringes at low $\rm{q_{z}}$ between the reflection edge and the $\rm{1^{st}}$ superlattice Bragg peak, testify for the uniform thickness of the entire film and a small surface roughness. The position of the superlattice Bragg-peaks are marked by arrows. As is expected for a superlattice with nearly equally thick LSCO and LCMO layers, the odd-order superlattice Bragg peaks ($\rm{1^{st}}$ and $\rm{3^{rd}}$ order) are very pronounced, whereas the even-order Bragg peaks ($\rm{2^{nd}}$ order peak) are essentially absent. The narrow width of the SL Bragg peaks is emblematic of a negligible variation of total thickness among the different BL repetitions and of its uniformity in the lateral direction. 
The model fit to this reflectivity curve (shown by the solid line) yields indeed very comparable thicknesses of  $9.8(1)$ nm and $9.6(1)$ nm for the LSCO and LCMO layers, respectively. These values are in excellent agreement with the estimates from the RHEED data. The rather low value of the average roughness of the LSCO/LCMO interfaces of $0.85(5)$ nm testifies for the high quality of the superlattice and it also compares well with the estimate from the AFM surface map of the 9\_BL sample that was shown in section \ref{AFM_sec}.

\begin{table}
\centering
\vspace*{0cm}
\caption{\label{Fitresult} Values of the parameters obtained from the fit to the unpolarized neutron reflectometry data in Fig. \ref{Neutron}.}
\begin{ruledtabular}
\begin{tabular}{l r}
\textit{$\rm{d^{LSCO}}$} & 9.8 $\pm$ 0.1 nm\\
\textit{$\rm{d^{LCMO}}$} & 9.6 $\pm$ 0.1 nm\\
$\rm{\sigma^{LSCO,LCMO}}$ & 0.85 $\pm$ 0.05 nm\\
\end{tabular}
\end{ruledtabular}
\end{table}
\subsection{\textbf{STEM}}
\label{STEM}

\begin{figure}[h!]
\vspace*{0cm}
\centering
\includegraphics[scale=1]{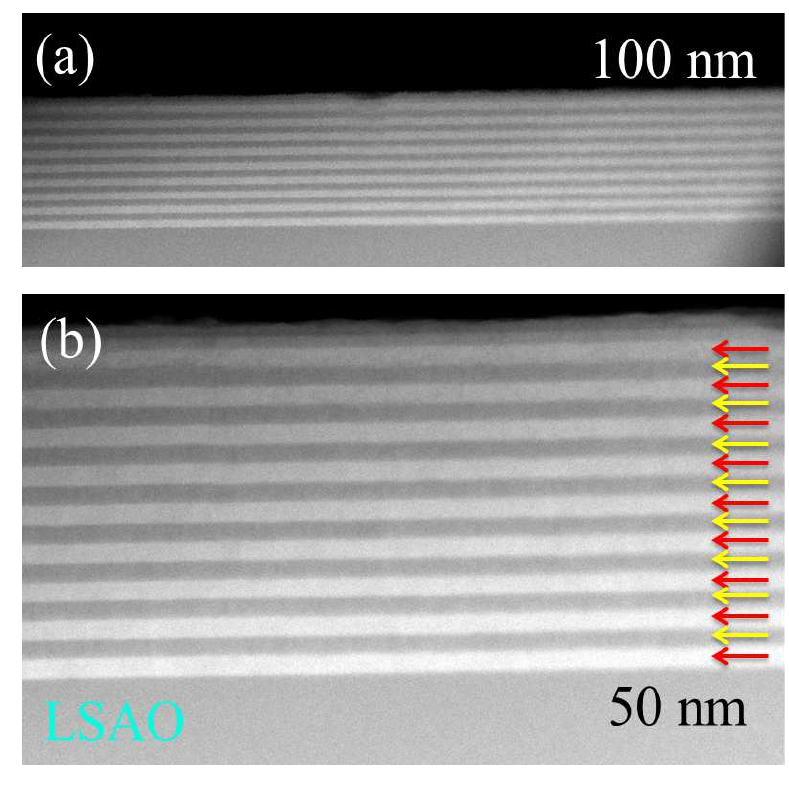}
\vspace*{0cm}
\caption{\label{LowMag} (Color online) (a) Low-magnification, high angle ADF images of a  [LSCO (7.5 u.c.)/LCMO (23 u.c.)]$\rm{_{9}}$ SL superlattice on a LSAO substrate. (b) Magnified image of the same sample. Red (yellow) arrows mark the LSCO(LCMO) layers.}
\end{figure}

In this paragraph we present the result of the scanning transmission electron microscopy measurements. Figure \ref{LowMag} shows low-magnification angular dark field (ADF) images of a LSCO/LCMO superlattice on a LSAO substrate. They confirm that the LSCO layers (bright layers, marked with red arrows) and the LCMO layers (dark layers, marked with yellow arrows) are continuous, coherent, and very flat over long lateral distances. They also show no indication of major defects or secondary phases.

\begin{figure*}
\vspace*{0cm}

\centering
\includegraphics[height=10cm,width=12cm]{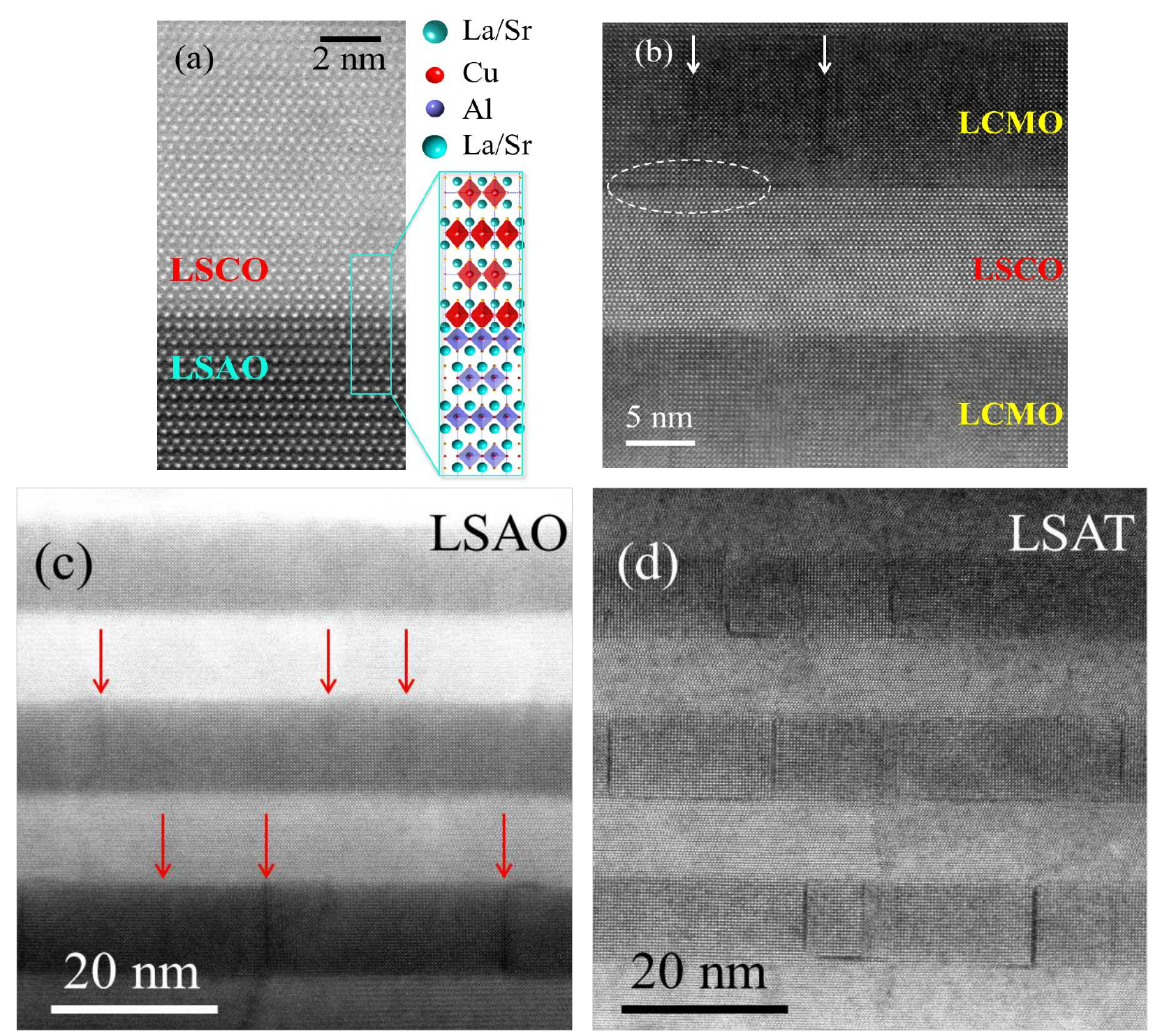} 
\vspace*{0cm}
\caption{\label{LSCO_LCMOinterface} (Color online) (a) Atomic resolution, high angle ADF image of the LSCO/LSAO interface. The sketch on the right shows the proposed interface structure (magnified scale) in the area that is highlighted with a rectangle. (b) Atomic resolution image of a LCMO-LSCO-LCMO stacking near the center of the superlattice. (c) Intermediate magnification image of the LCMO/LSAO superlattice and the LSAO substrate. Red arrows mark the position of antiphase boundaries in the LCMO layers. (d) Corresponding image of a LSCO/LCMO superlattice grown on a LSAT substrate. Antiphase boundaries are clearly visible as dark, vertical lines within the LCMO layers.}
\end{figure*}

Figure \ref{LSCO_LCMOinterface}(a) shows a high resolution Z-contrast image which depicts the LSAO/LSCO interface between the substrate and the $\rm{1^{st}}$ LSCO layer. The interface is coherent and the growth of the LSCO layer is epitaxial. The sketch shows the corresponding atomic stacking at the interface with apex-sharing $\rm{AlO_{6}}$ octahedra and $\rm{CuO_{6}}$ octahedra. Our proposed structural model suggests that LaO and SrO are removed from the surface of the (initially mix-terminated) substrate. This could happen either during the heat treatment of the substrate prior to the growth, or else right after the onset of the growth. This process may be reflected in the anomaly of the RHEED oscillation that has been observed for the $\rm{1^{st}}$ LSCO layer as was discussed in section \ref{RHEED_sec} and shown in Fig. \ref{RHEED}(e).

Figure \ref{LSCO_LCMOinterface}(b) displays a high magnification image of a LCMO-LSCO-LCMO sequence near the middle of the superlattice. The growth remains epitaxial and both the LCMO/LSCO and the interfaces are coherent and of high structural quality. Occasional defects are observed along the interface. This is visible on the upper left LCMO/LSCO interface (highlighted with a dashed white ellipse), where the LCMO and LSCO layers are shifted by half a LCMO unit cell. These mismatch-induced defects tend to be associated with antiphase boundaries (marked by arrows) which occur mostly in the LCMO layers at relative distances of about 20-50 nm. Figure \ref{LSCO_LCMOinterface}(c) displays an image of slightly lower magnification that shows the distribution of these antiphase boundaries which appear as vertical dark stripes (marked by the arrows). For comparison, Fig.  \ref{LSCO_LCMOinterface}(d) shows an image for a similar LSCO/LCMO SL grown on a $\rm{Sr_{0.7}La_{0.3}Al_{0.65}Ta_{0.35}O_{3}}$ (LSAT) substrate. It exhibits the same kind of vertical dark stripes due to the antiphase boundaries in the LCMO layers. This finding suggests that these extended defects are not caused by strain effects imposed by the substrate (since these differ strongly between LSAO and LSAT), but instead are caused by the lattice mismatch between the LCMO and the LSCO layers. More details about the analysis of the STEM data and of an element-specific electron energy loss spectroscopy study to reveal the atomic layer stacking at the LSCO-LCMO and LCMO-LSCO interfaces will be presented in a forthcoming publication \cite{Biskup}.

\subsection{Transport and Magnetization}
\begin{figure}
\centering
\vspace*{0cm}
\includegraphics[scale=.9]{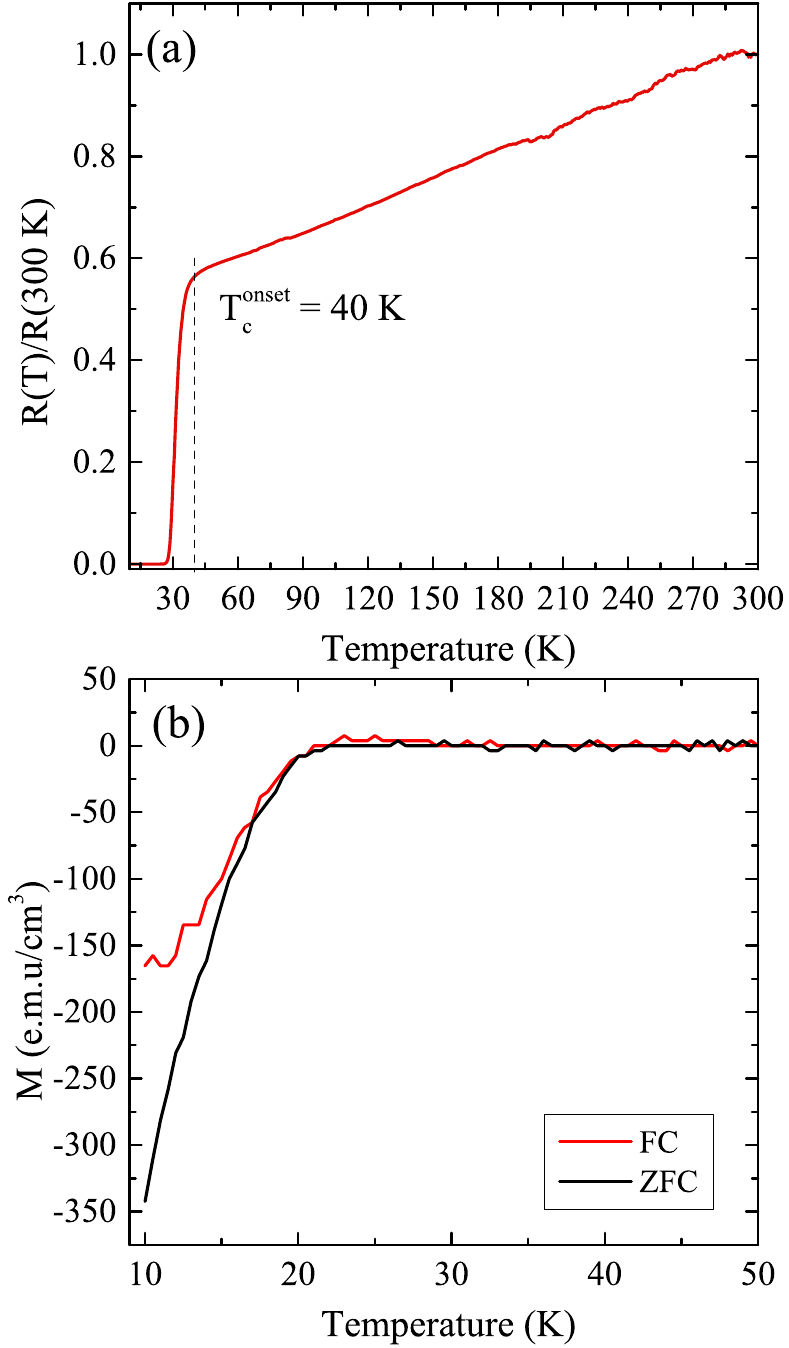}
\vspace*{0cm}
\caption{\label{LSCO_RMT} (Color online) (a) Temperature dependence of the resistance of the 10 nm thick LSCO film normalized to the value at 300 K. (b) Magnetization obtained in zero-field-cooled (ZFC) and field-cooled (FC) mode with a magnetic field of 200 Oe applied parallel to the c-axis.}
\end{figure}

Figure \ref{LSCO_RMT}(a) shows the temperature-dependent resistance normalized to the value at 300 K for the single LSCO (7.5 u.c.) thin film. The linear temperature dependence in the normal state is typical for an optimally doped cuprate high $\rm{T_{c}}$ superconductor. The onset temperature of the superconducting transition of $\rm{T^{onset}_{c}\approx40}$ K is also characteristic for optimally doped LSCO. The resistive transition is considerably broader than in high-quality bulk samples or in thicker LSCO films, i.e. the resistance vanishes (within the accuracy limits) only at $\rm{T_{c}(R\rightarrow0)\approx 25}$ K. We suspect that this broadening is caused by a certain amount of oxygen vacancies that are known to be crucial in obtaining a sharp resistive transition. We found that the transition width could not be significantly reduced by a postannealing treatment in flowing oxygen atmosphere at temperatures even upto 650 $\rm{\degree C}$ for a period of 30 hours. However, this failure to remove the oxygen defects may be due to the low oxygen diffusion rate along the c-axis of LSCO \cite{Claus1996}. Another factor might be a reduced concentration of the Sr cations, as compared to the target material, which could occur in PLD-grown films. However, this possibility is not supported by Rutherford back scattering (RBS) measurements that have been performed on a thicker LSCO film. The measurement yields a ratio of the cation stoichiometry of $\rm{La:Sr:Cu \approx 1.85:0.15:0.97}$.
To confirm that the superconducting state of the thin LSCO film is a macroscopic phenomenon, we performed magnetization measurements in zero-field-cooled (ZFC) and field-cooled (FC) modes with an external field of 200 Oe applied parallel to the c-axis of LSCO. A diamagnetic signal which sets in rather gradually and become prominent below about 20 K is seen in both the ZFC and FC curves in Fig. \ref{LSCO_RMT}(b). It develops rather gradually such that it is difficult to determine the onset temperature very accurately. Nevertheless, its onset appears to be reasonably close to the one of zero resistance at 25 K (see Fig. \ref{LSCO_RMT}(a)). Therefore, while the broadening of the superconducting transition may be characteristic of a certain degree of disorder, the sizeable diamagnetic response below 20 K is indicative of a bulk superconducting response. 

\begin{figure}
\centering
\vspace*{0cm}
\includegraphics[scale=1]{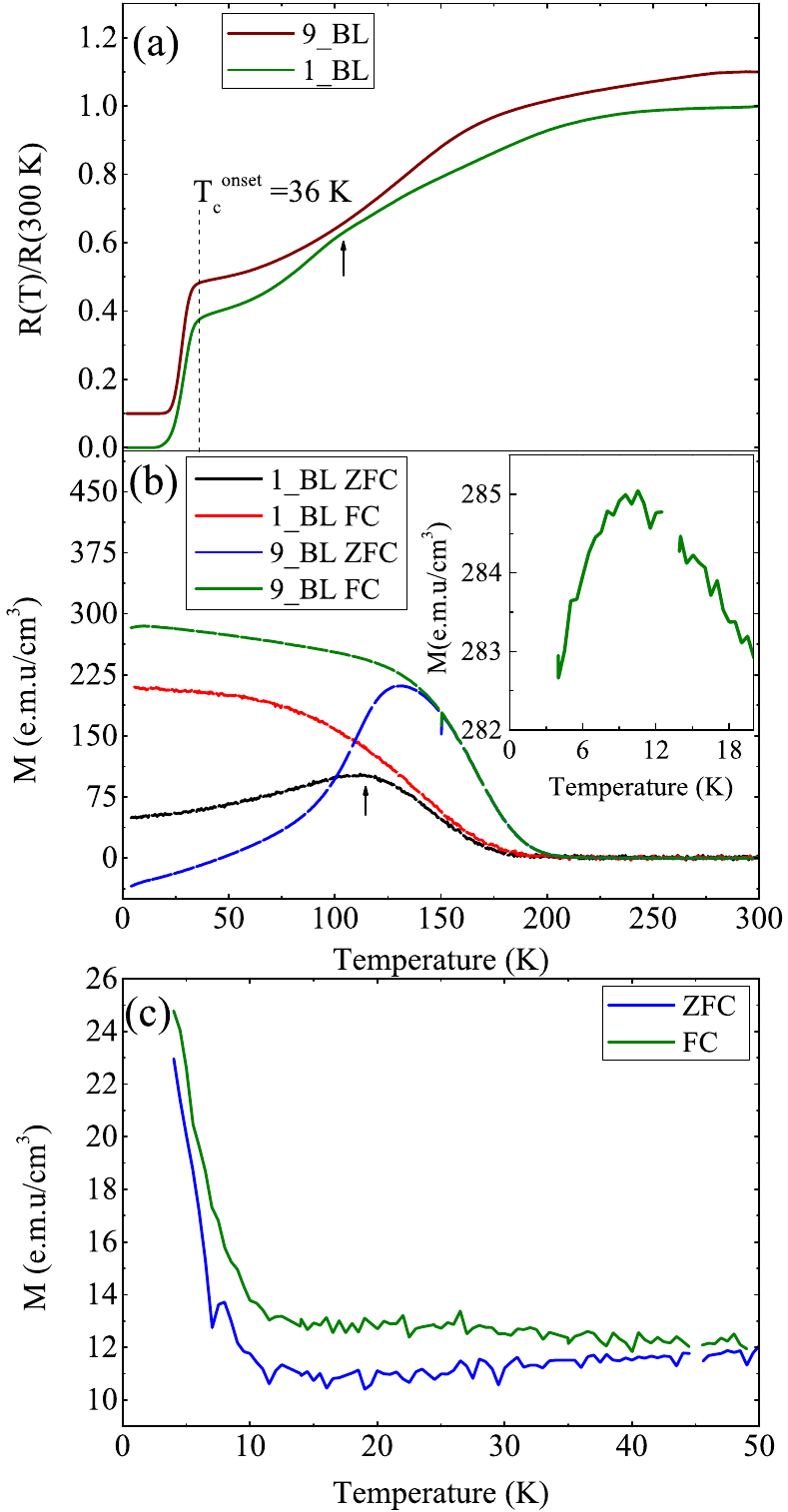}
\vspace*{0cm}
\caption{\label{SL_RMT} (Color online) (a) Resistance versus temperature curves of the 1\_BL (olive line) and 9\_BL (brown line) samples. The curves are vertically offset for clarity. (b) Magnetization curves measured in field-cooled (FC) and zero-field-cooled (ZFC) modes with a field of 100 Oe applied parallel to the layers. The inset shows a magnified view of the FC data of the 9\_BL sample at low temperature, it shows the weak diamagnetic signal which occurs below about 10 K. (c) Corresponding FC and ZFC magnetization curves with a magnetic field of 100 Oe applied perpendicular to the layers.}
\end{figure}

Figure \ref{SL_RMT}(a) shows the temperature dependent resistance curves (normalized to the value at 300 K) for the 1\_BL and 9\_BL LSCO (7.5-8 u.c.)/LCMO (23 u.c.) samples. Both samples show an onset of the superconducting transition at $\rm{T^{onset}_{c}\approx36}$ K and zero resistance at $\rm{T_{c}(R\rightarrow0)\approx15-20}$ K. Such a modest reduction of $\rm{T^{onset}_{c}}$ and $\rm{T_{c}(R\rightarrow0)}$, as compared to the LSCO thin film, may originate from the pair-breaking effect due to the proximity-coupling with the ferromagnetic LCMO layers. An alternative explanation is in terms of strain effects that can strongly influence the $\rm{T_{c}}$ values of LSCO films \cite{Sato2008}. The normal state resistance curves exhibit a clear change of slope around 180-190 K, a characteristic signature of the metal-to-insulator transition (MIT) of the LCMO layers. This conjecture is supported by the magnetization data in Fig. \ref{SL_RMT}(b) which show that a strong ferromagnetic signal starts to occur below  $\rm{T^{Curie}\approx190}$ K. In single LCMO films and in bulk LCMO this combined ferromagnetic and MIT transition is a well established feature that can be understood in terms of a competition between a Jahn-Teller effect, which localizes the charge carriers since it couples them to the local lattice distortions, and the ferromagnetic double-exchange interaction which requires itinerant charge carriers. The latter depends on the width of the conduction band and is therefore rather sensitive to the Mn-O-Mn bond angle \cite{Millis1998,Tokura1995}. This transition thus can be strongly affected by the strain  imposed by the substrate and/or the neighboring layers. The large compressive strain in the LSCO/LCMO SLs, which gives rise to a sizeable distortion of the $\rm{MnO_{6}}$ octahedra and a reduction of the Mn-O-Mn bond angles, is expected to result in a sizeable suppression of the MIT and of $\rm{T^{Curie}}$. The observed value of $\rm{T^{Curie}\approx190}$ K is indeed considerably lower than in bulk LCMO with $\rm{T^{Curie}\approx270}$ K \cite{Snyder1996} or in YBCO/LCMO SLs with  $\rm{T^{Curie}\approx225}$ K \cite{Malik2012} with less strained LCMO layers. A cation deficiency as the source of the suppressed $\rm{T^{Curie}}$ value is once more not supported by the RBS measurement. The obtained cation ratio of $\rm{La:Ca:Mn\approx 0.7:0.3:1}$ puts tight limits on a deviation of the stoichiometry from the ones of the $\rm{La_{2/3}Ca_{1/3}MnO_{3}}$ target. 

\begin{figure}
\centering
\includegraphics[scale=1]{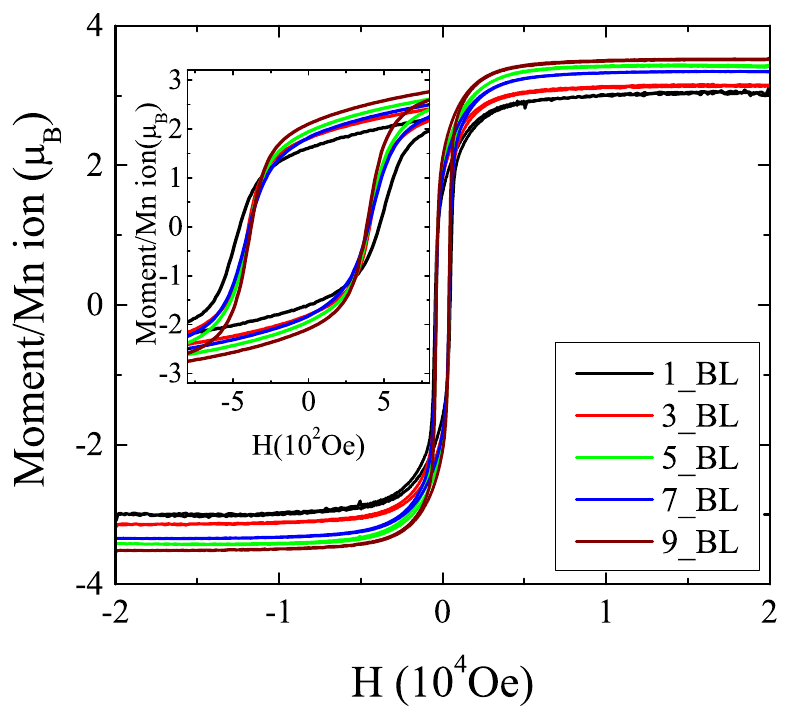}
\vspace*{0cm}
\caption{\label{SL_MH} (Color online) Magnetization-field (M-H) loops of the LSCO/LCMO SLs obtained at T=10 K after cooling in zero field. Inset: Magnified view of the hysteretic part of the M-H-loops.}
\end{figure}

The 1\_BL sample exhibits an additional kink (indicated by an arrow) in the resistance curve around 110 K. We suspect that this feature results from the enhancement of the compressive strain near the interface to the LSCO layer whose in-plane lattice parameter is locked to the one of the underlying LSAO substrate. While the field-cooled (FC) magnetization data of the 1\_BL sample in Fig. \ref{SL_RMT}(b) do not show a clear anomaly around 110 K, a pronounced peak is seen in the zero-field-cooled (ZFC) data (indicated by an arrow). A similar peak around 110-130 K along with a bifurcation of the FC and ZFC curves occurs for the other SLs. It highlights a hysteretic behavior that is also evident in the magnetisation loops at 10 K in Fig. \ref{SL_MH} which reveal a coercive field of about 460 - 380 Oe that is significantly larger than the applied field of 100 Oe for the data in Fig. \ref{SL_RMT}(b).

The magnetization data in the superconducting state exhibit only a very weak diamagnetic response that develops below about 10 K. This is shown in the inset of Fig. \ref{SL_RMT}(b) which displays on a magnified scale the FC magnetization data at low temperature for the applied magnetic field parallel to the layers. A similar effect occurs in the ZFC curve (not shown).  Figure \ref{SL_RMT}(c) shows the corresponding FC magnetization curve of the 9\_BL sample for the perpendicular magnetic field orientation (along the c-axis of LSCO). Instead of the expected diamagnetic response, the magnetization exhibits a rather sizeable increase below 10 K. A similar superconductivity-induced enhancement of the magnetization density was previously reported for YBCO/LCMO SLs and was  interpreted in terms of a magnetic flux compression due to non-equilibrium vortex pinning effects \cite{Lopez2006,Chen2010}. The explanation of this effect is debated and beyond the scope of this manuscript. Another anomalous feature that is presently not well understood concerns the sizeable difference between the onset temperatures of the paramagnetic signal (and the weak diamagnetic signal for H $\parallel$ ab) of about 10 K and of the superconducting transition in the resistance with an onset at 36 K and zero resistance around 20 K. A similar difference was observed in YBCO/LCMO superlattices \cite{Malik2012}. It may be caused by a sizeable inhomogeneity of the superconducting transition within the LSCO layers, but it may also be 
an intrinsic feature, for example, due to the formation of a spontaneous vortex lattice phase \cite{Bernhard2000}. Irrespective of these open questions, we can conclude that below 10 K these LSCO/LCMO superlattices exhibit a bulk superconducting response. The nature of the superconducting state at intermediate temperatures will require further investigations, for example with magnetic scanning probe or small angle neutron scattering techniques which can directly probe the  magnetic vortex lattice. 

The saturation value of the Mn moments in the LCMO layers has been determined from the magnetization loops  measured at 10 K with the magnetic field parallel to the layers. Figure \ref{SL_MH} shows that the saturation magnetic moment increase rather gradually from about 3.0(1) $\rm{\mu_{B}}$ per Mn ion for the 1\_BL sample to about 3.5(1) $\rm{\mu_{B}}$ per Mn ion for the 9\_BL sample. The latter is very close to the maximal value of  3.67 $\rm{\mu_{B}}$ in bulk LCMO. This confirms that all the LCMO layers exhibit a predominant ferromagnetic order. The inset shows the evolution of the coercive field,  $\rm{H_{c}}$, which exhibits a small, yet marked difference between the sample with 1 BL and the ones with 3-9 BLs. The value for the 1\_BL sample of  $\rm{H_{c}\approx 460}$ Oe is enhanced as compared to $\rm{H_{c}\approx 380-400}$  Oe in the 3-9\_BL samples. A similar enhancement of $\rm{H_{c}}$ has recently been reported for $\rm{La_{2}CuO_{4}/La_{0.7}Ca_{0.3}MnO_{3}}$ bilayers. There it has been interpreted in terms of the frustration of the Mn spins due to the exchange coupling across the interface with the antiferromagnetic Cu moments \cite{Ding2013}. However such a static antiferromagnetic order of the Cu moment is not present in our LSCO/LCMO bilayers which contain optimally doped LSCO layers. This leads us to suggest an alternative interpretation in terms of an enhanced pinning of the ferromagnetic domains due to the strain induced disorder that is most pronounced for the $\rm{1^{st}}$ LCMO layer. 
\section{Summary}
In summary, we have reported the pulsed laser deposition growth of thin films of  $\rm{La_{1.85}Sr_{0.15}CuO_{4}}$ and $\rm{La_{1.85}Sr_{0.15}CuO_{4}/La_{2/3}Ca_{1/3}MnO_{3}}$  superlattices on LSAO substrates. We also characterized their structural, electronic and magnetic properties. In particular, we have shown that these superlattices can be grown epitaxially and with a high structural quality. The strain relaxation has been analyzed for a series of samples for which the number of repetitions of the LSCO/LCMO bilayers increases from X = 1 to 9. It was found that the first LSCO layers remains clamped to the substrate whereas the first LCMO layers exhibits a sizeable strain relaxation. The in-plane parameters of the subsequent LSCO and LCMO layers alternate around an average value that remains almost constant such that the former are under a sizeable tensile strain and the latter under a corresponding compressive strain. The LSCO layers accommodate these strain effects without forming any major structural defects, in contrast, a sizeable number of vertical antiphase boundaries is observed in the LCMO layers. Despite these strain effects, we have shown that the LSCO layers remain superconducting with a relatively high onset temperature $\rm{T^{onset}_{c}}\approx 36$ K and also with a noticeable response in the magnetization data. In the latter we observe a weak diamagnetic signal for the magnetic field parallel to the layers and a rather large and anomalous paramagnetic response for the perpendicular field direction. The latter can be explained in terms of a flux compression due to an anomalous vortex pinning effect that has been previously observed in corresponding YBCO/LCMO superlattices. Our results confirm that these LSCO/LCMO superlattices can be used as a second model system (besides the YBCO/LCMO superlattices) that allows one to study the proximity coupling between the  ferromagnetic and superconducting orders as to identify the underlying mechanism. In particular, the LSCO/LCMO system allows for a systematic change of the hole doping state of the cuprate layers which can be varied over the entire superconducting part of the phase diagram (and beyond).

\begin{acknowledgments}
The research in Fribourg was supported by the Schweizer Nationalfonds (SNF) grant 200020-140225 and the National Centre of Competence in Research ``Materials with Novel Electronic Properties - MaNEP". Research at ORNL (MV) was supported by the U.S. Department of Energy (DOE), Basic Energy Sciences (BES), Materials Sciences and Engineering Division. Neven Biskup was supported by the ERC starting Investigator Award, grant $239739$ STEMOX.  Electron microscopy observations carried out at the ICTS- Centro Nacional de Microscopia Electronica (UCM). The experiment at NREX has been supported by the European Commission under the 7th Framework Programme through the  `Research Infrastructures' action of the `Capacities' Programme, NMI3-II Grant number 283883.

\end{acknowledgments}

\end{document}